\documentstyle[11pt]{article}
\input{epsf}

\newcommand{\n}{\noindent}

\def\12{{1\over 2}}
\def\msun{{M_\odot}}
\def\be{\begin{equation}}
\def\ee{\end{equation}}
\def\ba{\begin{eqnarray}}
\def\ea{\end{eqnarray}}
\def\de{\partial}

\def\ltsima{$\; \buildrel < \over \sim \;$}
\def\simlt{\lower.5ex\hbox{\ltsima}}
\def\gtsima{$\; \buildrel > \over \sim \;$}
\def\simgt{\lower.5ex\hbox{\gtsima}}

\begin{document}
\title{\bf The generation of low-energy cosmic rays in molecular clouds
\footnote{Astr Rept: received Sep 10, 2003; accepted Sep 20, 2004}}
\author{Yuri A. Shchekinov\thanks{yus@phys.rsu.ru}\\
\\
        Dept of Physics, Univ of Rostov,
        Rostov on Don, Russia and\\
        Astron Inst, Ruhr-Universit\"at Bochum,
        Bochum, Germany}

\date{}
\maketitle

\begin{abstract}
It is argued that if cosmic rays penetrate into molecular clouds, the total energy
they lose can exceed the energy from galactic supernovae shocks. It is shown
that most likely galactic cosmic rays interacting with the surface layers
of molecular clouds are efficiently reflected and do not penetrate into the cloud
interior. Low-energy cosmic rays ($E<1$ GeV) that provide the primary
ionization of the molecular cloud gas can be generated inside such clouds by
multiple shocks arising due to supersonic turbulence.

\end{abstract}

\section{Introduction}

Gas phase chemical reactions in molecular clouds are
predominantly ion-molecular reactions, and are catalyzed by
H$^+$ ,H$^+_2$, and H$^+_3$ ions, which are
produced, in turn, by low-energy cosmic rays ($E \simlt
1$ GeV; see the review [1]). The primary ionization
rate by cosmic rays required to sustain the fractional
ionization and concentrations of these ions is
estimated to be $\zeta\sim 10^{-17}$ s$^{-1}$. A similar value is
obtained for the upper limit to the primary ionization rate derived
from the abundances of HD in molecular
clouds, $\zeta\simlt 10^{-16} -10^{-17}$ s$^{-1}$ [2, 3]. It was first
proposed in [4] that the fractional ionization in the
cores of molecular clouds might exceed the value
corresponding to $\zeta\sim 10^{-17}$ s$^{-1}$ by a factor of five,
which would require that $\zeta$ be increased by a factor
of 25. Based on the assumption that the predominant magnetic field
structure in molecular clouds is dipolar, it was shown in [5] that
cosmic rays with energies $E>1$ MeV/nucleon essentially freely penetrate into
their interior, with ionization losses constituting about
10\% of their energy (see also [6]). For this reason,
the above estimates of the primary ionization rate are
considered to be valid for the interstellar medium as
well. However, new data on the properties of molecular clouds and new
theoretical analysis of their nature have recently cast doubts on the
unconditionality of the conclusions of [5] and the general view
that the primary ionization rates by cosmic rays in molecular clouds and
in the ISM are in agreement. We present here arguments supporting the idea
that cosmic rays with energies $E\simlt 1$ GeV are not able to penetrate into
the interiors of molecular clouds, so that the presence of such particles
there suggests that high-energy, nonrelativistic, nonthermal particles
are generated within the molecular clouds themselves. This could occur,
in particular, due to MHD shock waves arising in clouds during their
formation from diffuse gas.

Section 2 presents estimates of the total energy lost by galactic
cosmic rays in molecular clouds, and
shows that these energy losses exceed the total energy rate
injected by supernovae into cosmic rays. Section 3
presents arguments based on the results of recent
numerical modeling that suggest that the magnetic
field structure in molecular clouds should differ from a
simple dipolar field, and probably has a toroidal (helical)
component in the plane perpendicular to the rotational axis of the cloud.
In this case, most of the cosmic rays incident on the cloud should be
reflected. Section 4 presents estimates of the efficiency of the
acceleration of cosmic rays by MHD shock waves in
molecular clouds. Finally, Section 5 summarizes our
conclusions.

\section{Energy losses of cosmic rays in molecular clouds}

It can easily be shown that the dominant contribution to the
primary ionization rate by galactic cosmic rays with an energy
spectrum [7]

\be
J(E)\simeq {3\times 10^4 \over
(800 + E)^{2.5}}~{\rm (cm^2~sr~s~MeV)^{-1}},~E<10~{\rm GeV}
\ee
is provided by cosmic rays with energies $E<1$ GeV (the energy $E$ in (1)
is in MeV); the spectrum has a break at energies $E>10$ GeV, where its slope
becomes steeper. The fraction of the total energy $E = \gamma m_p c^2$ lost
by a cosmic ray particle (proton) as it passes through material with a
hydrogen density  along the path $N_L$ is determined to order of magnitude as
[6]

\be
\delta_E = {4\pi e^4\Lambda\over  m_e c^2 m_p c^2}
{\gamma\over \gamma^2-1} N_L \simeq (0.5 - 1)\times 10^{-4}
{\gamma\over \gamma^2-1} N_{L,22},
\ee
where $\Lambda=\ln[2m_e c^2 \beta^2 /I(1 - \beta^2)] - \beta^2 \simeq 5 - 12$
for energies from 1 MeV to 1 GeV, $N_{L,22} =10^{-22} N_L$,
$\beta= v/c$, and $c$ is the speed of light. As it penetrates into a cloud,
a high-energy particle undergoes numerous scatterings, so that the total path
length travelled by the particle in the cloud is $L\sim 4cR^2/D$, where $R$
is the radius of the cloud and $D$ is the diffusion coefficient for cosmic
rays in the cloud. The total density of particles along the trajectory is
$N_L\sim 2NR/\lambda\gg N$ [6], where $\lambda$ is the characteristic
correlation scale for turbulence in the magnetic field, $\lambda\ll R$, and
$N$ is the surface density of the molecular cloud. Observed molecular clouds
display a roughly constant (independent of radius) surface density of
$N =1.5\times 10^{22}$ cm$^{-2}$ (see the review [8]). Thus,

\be
\delta_E\sim (1.5-3)\times 10^{-4}{\gamma \over
\lambda_R (\gamma^2 - 1)},
\ee
$\lambda_R=\lambda/R$; for $E_k=100$ MeV and $\lambda_R=0.1$, we find
$\delta_E \sim 0.01$.

The energy absorbed by all the molecular clouds in the Galaxy is

\be
\dot E^{cr}_{ MC} = S \int\limits_{100 ~{\rm MeV}}^\infty
\delta_E EJ(E)dE\sim 4\times 10^{-5} S\lambda_R^{-1} ~{\rm erg/s},
\ee
where $S$ is the total surface area of a molecular cloud. Adopting for the
entire Galaxy the molecular cloud spectrum obtained in [9],

\be
{d{\cal N }\over d M}=AM^{-1.73},
\ee
over the interval $M=[M_1=10^2,M_2=10^6]\msun$ with the normalization factor
$A=3.5\times 10^7$, we can find the total surface area of clouds

\be
S=4\pi\int\limits_{M_1}^{M_2}R^2 {d{\cal N}\over dM}
dM =7\times 10^{46}\sigma n^{-2/3},
\ee
where $n\sim 10^2$ cm$^{-3}$ is the mean gas density in the
clouds, and $\sigma>1$ is a factor accounting the fact that the
surface area of a cloud is greater than that of a sphere due to irregularities
in the cloud boundary [10]. When calculating the normalization coefficient
$A$, we assumed that the mass spectrum of molecular clouds determined in [9]
for the Perseus arm is valid for the Galaxy as a whole, with the mass
of molecular gas in the Galaxy being $5\times 10^9\msun$. This yields
$S \simeq 3 \times 10^{45}\sigma$ cm$^2$, and $\dot E^{cr}_{MC}\simeq 2 \times
10^{42}\sigma/\lambda_R$ erg s$^{-1}$ for the energy losses of cosmic rays in
clouds. The production of energy in galactic cosmic rays in supernovae is
$\dot E^{cr}_{SN} \sim 10^{42}\eta\nu_{1/30}$ erg s$^{-1}$, where
$\eta<1$ is the fraction of the supernova energy that
goes into cosmic rays (for which modern estimates yield $\eta\sim 0.1$
[11]), and $\nu_{1/30}$ is the supernova rate in the Galaxy in units of
explosions per 30 years. It can readily be verified that, for $\eta\sim 0.1$
and $\lambda_R \sim 0.1$, the amount of energy in cosmic rays that is
absorbed by molecular clouds is higher than the amount of cosmic-ray energy
produced in supernovae. In fact, this conclusion can be stated more firmly:
with the adopted assumptions, $\dot E^{cr}_{SN} \ll \dot E^{cr}_{MC}$,
which implies that probably $\lambda_R \ll 0.1$. Indeed, if in reality
the correlation length is determined by viscosity, then $\lambda_R \sim
(\ell/R)^{3/4}$, where $\ell$ is the mean free path of the particles
($\ell\ll  R$) [6]. Recent numerical modeling of MHD turbulence
in molecular clouds [12, 13] suggests that the correlation length is
0.01-0.03 of the scale on which energy is injected into the system,
which in our case is the radius of the cloud. To resolve
this inconsistency between the cosmic ray energy that could be absorbed
in molecular clouds and the cosmic ray energy produced by supernovae, we must
suppose that galactic cosmic rays interacting with molecular clouds are
predominantly reflected, so that only a small fraction of the cosmic ray
energy is lost.

\section{Reflection of cosmic rays from the surface layers of molecular clouds}

According to recent numerical modeling, molecular clouds represent transient
regions of enhanced density arising due to the action of converging flows of
interstellar gas [14--17], whose lifetimes, $t_{MC}\sim 2\times 10^7
-10^8$ yr, are probably determined by the activity of the stars born in
them. In this picture, one of the main sources of turbulent energy
in molecular clouds is the kinetic energy of the converging flows, although
it is not ruled out that an appreciable amount of energy can be contributed
by young stars born in the molecular cloud cores. During the formation of
a molecular cloud -- i.e., the transformation of atomic into molecular
hydrogen in the process gas compression -- some fraction of its rotational
angular momentum is lost, but the remaining specific angular momentum of the
cloud is still substantial: on average, the specific angular
momentum in molecular clouds is a factor of four lower than the
specific angular momentum of the diffuse interstellar gas. This means that
on the lifetime a cloud rotates about 30 times as it is compressed in the
converging flows (we have adopted here a characteristic rotational velocity
for the cloud $\Omega\sim 10^{-14}$ s$^{-1}$ [18]). Thus, we expect
the magnetic field in the outer parts of the molecular cloud to have
a helical structure with an appreciable toroidal component (i.e., the
component perpendicular to the rotational axis of the cloud). This type of
field structure is indeed observed in a number of cases (such as the
cloud Lynds 1641) [19]. In this case, no matter what the angle at which a
cosmic ray particle is incident on the molecular cloud surface, the pitch
angle will reach its critical value as the particle penetrates
into the cloud, causing the cosmic ray particle to experience mirroring.
The characteristic time for the reflection of cosmic rays by this process
is [5]

\be
t_m \sim {\gamma m_p \over p}\left({\de \ln B\over \de s}\right)^{-1}
\sim {\gamma m_p L_B \over p},
\ee
where $L_B$ is the length along the trajectory of the cosmic ray particle
over which the regular component of the magnetic field varies, for which it
is natural to adopt $L_B \sim R$. The characteristic diffusion time for
the cosmic rays is

\be
t_D \sim {R^2 \over D}.
\ee
If we suppose that this diffusion is determined by kinetic processes with
cross section $\sigma_i \sim 3 \times 10^{-18}$ cm$^2$ (the cross section for
ionization losses), the diffusion time will be $t_D =(N\sigma_i)^{3/4}R/c$,
so that

\be
{t_m \over t_D}\sim {1 \over (N\sigma_i)^{3/4}} \ll 1.
\ee
If the diffusion is determined by resonance scattering on small scales of the
order of the gyroradius of a cosmic ray proton, $l \sim 2\pi r_p$ [5, 6], the
diffusion coefficient will be $D \simlt cR^{1/4}l^{3/4}$, and the ratio of
the time scales becomes

\be
{t_m \over t_D} \sim \left({l\over R}\right)^{3/4}\ll 1.
\ee
Thus, cosmic rays should be reflected by the enhanced magnetic field of
a molecular cloud appreciably more rapidly than they can penetrate into the
cloud interior. The fraction of cosmic rays that are able to penetrate into
the cloud can be crudely estimated using the ratio $t_m/t_D$, which is to
order of magnitude $10^{-3}$ in both cases. This analysis forces us to
conclude that the primary ionization rate of the molecular gas is substantially
lower than the value that is required in order for the gas--chemical reactions
in the cloud to be sustained: $\zeta \sim 10^{-17}$ s$^{-1}$. Indeed, galactic
cosmic rays with the spectrum presented above provide to order of
magnitude just this primary ionization rate. Even if about 10\% of the
high-energy particles can penetrate into the cloud, which could
with some margin lead to agreement between $\dot E^{cr}_{MC}$ and $\dot
E^{cr}_{SN}$, $\zeta$ would be at least an order of magnitude lower than the
required value. This contradiction can be resolved assuming that turbulent
motions in the molecular clouds are able to accelerate nonthermal particles to
sufficiently high energies to sustain the required level of primary ionization.

\section{Acceleration of nonthermal particles in molecular clouds}

Molecular clouds display well-developed turbulence, as a rule, supersonic, and
apparently super-Alfv\'enic (see the discussions in [4, 12, 20]). It is,
accordingly, natural to expect the presence of numerous MHD shock waves in the
cloud, as is observed in both numerical simulations (see the discussion in [21])
and in real molecular clouds [22, 23]. The characteristic velocities of MHD
shocks in molecular clouds can reach $\sim 10$ km s$^{-1}$, in which case it
is reasonable to expect that some fraction of their kinetic energy could
be transformed into nonthermal particles accelerated in the shock fronts. The
efficiency of particle acceleration under these conditions can be estimated
using the theory developed in the studies [24--27], which are concerned with
the acceleration of particles in the fronts of chaotic shocks in turbulent
interplanetary and interstellar fields. In this theory, the maximum energy
that can be gained by particles is determined by the ratio of the time over
which the acceleration region exists (the time for the shocks to travel
through the distance separating them), $t_a\sim L_s/u$, and the time for
the diffusion of the particles from the acceleration region,
$t_D\sim L_s^2/D$, where $L_s$ is the characteristic distance between the
shock fronts and $u$ is the shock velocity [26] (the accelerated particles
leave the acceleration zone when $t_D/t_a<1$). The diffusion coefficient
$D=v\Lambda/3$ is determined by the scattering of particles on small-scale
inhomogeneities in the magnetic field with a characteristic transport free
path $\Lambda\simeq C_\nu L_s(r_p/L_s)^{2-\nu}$, where $C_\nu\simeq 0.3$,
$r_p=cp/eB$ is the gyroradius of the particle, $p$ is the momentum of the
particle, and $\nu$ is the index of the spectrum of the small-scale
fluctuations of the magnetic field ($d\langle B^2\rangle/dk\sim k^{-\nu}$,
$kL_s\gg 1$ [26, 27]). We will estimate the maximum energy of the accelerated
particles $E_M$ from the condition $t_D/t_a=1$. This yields for the spectral
index $\nu=1.5$

\be
E_M\sim 5\times 10^{-3}L_{s,R}^2 B_1^2 R_1^2~ {\rm GeV},
\ee
when $E_M<m_pc^2$ and

\be
E_M\sim 0.1L_{s,R}B_1R_1~{\rm GeV},
\ee
when $E_M>m_pc^2$; here, $L_{s,R}=L_s/R$, $R_1=R/(1~ {\rm pc})$,
$B_1=B/(1~\mu{\rm G})$. Thus, in clouds with characteristic parameters
$B \sim 10\mu$G and $R\sim 10$ pc and with $L_s/R\sim 0.01-0.03$, we expect
protons to be accelerated to energies $E_M\sim 5-15$ MeV. However,
$E_M$ is very sensitive to the index of the magnetic field fluctuation
spectrum $E_M\propto (u/c)^{1/(2-\nu)}$ in the relativistic limit and
$E_M\propto (u/c)^{2/(2-\nu)}$ in the nonrelativistic case. Therefore, with
a Kolmogorov turbulence spectrum, $\nu\simeq 1.7$, and shock velocities
$u\sim 10$ km s$^{-1}$, the estimates of $E_M$ fall to values of the
order of atomic energies. From this point of view only detailed observations of
the magnetic field fluctuation spectrum will enable confident conclusion of
whether or not nonthermal particles can be accelerated to sufficiently high
energies in molecular clouds. Recent numerical simulations of MHD turbulence
in molecular clouds suggest that, in most cases, a Kolmogorov spectrum is
established near wavenumbers $kL/2\pi \simgt 10$, but the spectrum becomes
appreciably non-Kolmogorov at longer wavelengths (exceeding 0.1 of the cloud
radius), where it has an index near zero [12, 13]. Thus, the transition from large
scales, corresponding to the energy injection scales, to the scales on which
an inertial regime is established encompasses a broad wavenumber range. If we
associate the possibility of generation of cosmic rays with long-wavelength
components of the turbulent motions in the molecular cloud, the energy $E_M$
could reach values of 1 GeV or higher. However, both the steady state
characteristic energy of the nonthermal particles $E_\ast$ and the spectrum of
the generated cosmic rays are determined to a considerable extent by the
injected particle distribution function -- i.e., by those particles that,
having entered in an active acceleration region, are able
to undergo fairly frequent interactions with the shocks there -- and by the
fraction $\eta$ of the injected particles in the flow as a whole that intersect the
shock front. One source of particles that is usually considered is the
particles that become extracted from the thermal background during the
scattering of fairly energetic protons on forming collisionless shocks. This
process is probably not efficient in molecular clouds, because the thermal
energy of the particles here is too low. Winds from young stars and/or
low-energy cosmic rays entrained in the converging flows forming the molecular
cloud may serve as a source of such particles. In any case, the question
remains open and requires a separate analysis; more detailed discussions in
application to shocks from supernovae and to the diffuse interstellar
and interplanetary media can be found in [24, 26, 28, 29]. The value of the
fraction of interacting injected particles $\eta$ is very uncertain. The
injection rate at the bow shock in the magnetosphere of the Earth is
$\eta\sim 10^{-3}$ [30]. Estimates carried out in [31] for supernova remnants
leave open a very wide range for $\eta$: $10^{-5}<\eta<10^{-1}$. If we
adopt the lower limit for molecular clouds, the characteristic energy of the
nonthermal particles proves to be too low. Based on the fact that the
characteristic energy of the nonthermal particles should be such that their
total energy $W_p\sim \eta np_\ast^2/2m_p$ does not exceed the kinetic
energy of the generated turbulent motions $W_T\sim \rho u^2/2$,
we arrive at the estimate $E_\ast\simlt 0.5\eta^{-1} m_pc^2(u/c)^2\sim
50$ keV. The Larmor radius of protons with such energies is only
$r_p\sim 3\times 10^{9}$ cm in a magnetic field of $B\sim 10\mu$G, so that
the protons should primarily occupy a limited volume near the surface of the
accelerating shock.

To estimate the primary ionization rate provided throughout the molecular
cloud by the generated nonthermal particles, we suppose
that only those particles that at times $t\sim t_a$ have diffusion lengths
equal to half the distance between the shock fronts (i.e., $\sim L_{s,R}=1/2$)
are able to occupy the cloud volume fairly uniformly and contribute to the
ionization of the medium in the cloud. Adopting a power-law spectrum for the
nonthermal particles, $dN_p/dE\propto E^{-q}$, and assuming that the total
energy contained in these particles is comparable to the energy of the
turbulent motions, $\int EdN_p\sim W_T$, we obtain the primary ionization
rate

\be
\zeta= \int\limits_{E_{M/2}}^{E_{M}} v(E) {dN_p\over dE}\sigma(E)dE
\simeq 3\times 10^{-18} n~{\rm s^{-1}},
\ee
for $q=2$ and

\be
\zeta\sim 3\times 10^{-19} n~{\rm s^{-1}},
\ee
for $q=2.5$; here, $n$ is the gas density in the molecular cloud. Thus, for
a mean gas density $n\sim 10^2$ cm$^{-3}$ [32], we find $\zeta\sim 10^{-17}-
10^{-16}$ s$^{-1}$, which is close to the value required to sustain the
gas-phase chemistry in the molecular clouds. Note that larger values exceed
the value expected from galactic cosmic rays. It is worth stressing in this
connection that recent observations of some molecular clouds require primary
ionization rates that cannot be provided by galactic cosmic rays. In particular,
the value $\zeta=5.6\times 10^{-17}$ s$^{-1}$ with an uncertainty factor of
about three is preferred in [33]; i.e., the upper end of
the allowed interval corresponds to $\zeta\sim 10^{-16}$ s$^{-1}$.
However, even the most probable value exceeds the value expected from galactic
cosmic rays by a factor of five. This can be considered as an additional argument
for the need for additional sources of cosmic rays in molecular clouds and the
possible generation of cosmic rays inside them.

\section{Conclusions}

We have shown the following:

1. The energy lost by galactic cosmic rays in molecular clouds may exceed the
total energy converted into cosmic rays by supernovae.

2. An appreciable fraction of cosmic rays with energies $E \sim 1$ GeV can
be reflected from the surface layers of molecular clouds through mirror
effect. Thus, an additional source of nonthermal particles is required to
sustain the gas-phase chemical processes occuring in these clouds.

3. One such source is particles accelerated on the fronts of MHD shocks that
arise during the formation of molecular clouds. The primary ionization
rate that can be provided by such particles is $\zeta\sim 10^{-17}-10^{-16}$
s$^{-1}$.

\section*{Acknowledgements}

The author thanks the referee for comments. This
work was supported by the Russian Foundation for
Basic Research (project no. 02-02-17642), the INTAS foundation
(project no. 99-1667), and the German Science Foundation (DFG;
project no. SFB 591, TP A6).This work has made use of the NASA Astrophysics
Data System Abstract Service.

\section*{References}

1. W. D. Watson, Rev. Mod. Phys. 48, 513 (1976).

\n
2. E. J. O'Donnell and W. D. Watson, Astrophys. J. 191,
89 (1974).

\n
3. J. Barsuhn and C. M. Walmsley, Astron. Astrophys.
54, 345 (1977).

\n
4. P. C. Myers and V. K. Khersonsky, Astrophys. J. 442,
186 (1995).

\n
5. C. J. Cesarsky and H. J. V\"olk, Astron. Astrophys. 70,
367 (1978).

\n
6 . T. Nakano and E. Tademaru, Astrophys. J. 173, 87
(1972).

\n
7. J. A. Simson, Ann. Rev. Nucl. Part. Sci. 33, 330 (1983).

\n
8. C. F. McKee, in: The Origin of Stars an Planetary
Systems, Ed. by C. J. Lada and N. D. Kylafis (Kluwer
Acad., 1999), p. 29.

\n
9. M. H. Heyer and S. Terebey, Astrophys. J. 502, 265 (1998).

\n
10. E. Falgarone and J.-L. Puget, Astron. Astrophys.
162, 235 (1986).

\n
11. R. Schlickeiser, Cosmic Ray Astrophysics (Springer,
Berlin, 2002).

\n
12. J. Cho and A. Lazarian, Phys. Rev. Letters 88, 245001
(2002).

\n
13. J. G. Vestuto, E. C. Ostriker, and J. M. Stone, Astrophys.
J. 590, 858 (2003).

\n
14. J. Ballesteros-Paredes, E. V\'azquez-Semadeni, and
J. Scalo, Astrophys. J. 515, 286 (1999).

\n
15. E. V\'azquez-Semadeni, in: New Perspectives in the
Interstellar Medium, Ed. by A. R. Taylor, T. L. Landecker,
and G. Joncas (Astron. Soc. Pacif., San Francisco, 1999),
ASP Conf. Ser. 168, p. 345.

\n
16. W.-T.Kim, E. C. Ostriker, and J. M. Stone, Astrophys. J. 581, 1080 (2002).

\n
17. R. S. Klessen, Rev. Mod. Astron. 16 (2003).

\n
18. J. P. Phillips, Astron. Astrophys., Suppl. Ser. 134, 241
(1999).

\n
19. F. J. Vrba, S. E. Strom, and K. M. Strom, Astron. J. 96, 680 (1988).

\n
20. S. Boldyrev, Astrophys. J. 569, 841 (2002).

\n
21. P. Padoan, A. A. Goodman, and M. Juvela, Astrophys. J. 588, 881 (2003).

\n
22. E. Falgarone, J.-L. Puget, and M. Perault, Astron. Astrophys. 257,
715 (1992).

\n
23. P. C. Myers and C. F. Gammie, Astrophys. J. 522,
L141 (1999).

\n
24. I. N. Toptygin, Cosmic Rays in Interplanetary Magnetic Fields
(Reidel, Dordrecht, 1985; Nauka, Moscow, 1983).

\n
25. A. M. Bykov and I. N. Toptygin, Zh. Eksp. Teor. Fiz.
98, 1255 (1990) [J. Exp. Theor. Phys. , (1990)].

\n
26. A. M. Bykov and G. D. Fleishman, Pis'ma Astron. Zh. 18, 234 (1992)
[Astron. Lett. 18, 95 (1992)].

\n
27. A. M. Bykov and G. D. Fleishman, Mon. Not. R. Astron. Soc. 255, 269 (1992).

\n
28. D. C. Ellison and D. Eichler, Astrophys. J. 286, 691
(1984).

\n
29. R. Blandford and D. Eichler, Phys. Rep. 154, 2 (1987).

\n
30. M. A. Lee, J. Geophys. Res. 87, 5063 (1982).

\n
31. L. O'C. Drury, W. J. Markiewicz, and H. J. V\"olk,
Astron. Astrophys. 225, 179 (1989).

\n
32. J. Bally, in: Galactic Structure, Stars an the Interstellar Medium. ASP
Conf. Ser, Eds. C. E. Woodward, M. D. Bicay, and J. M. Shull (Astron. Soc.
Pacif., San Francisco, 2001), ASP Conf. Ser. 231,
p. 204.

\n
33. S. D. Doty, E. F. van Dishoeck, F. F. S. van der Tak,
and A. M. S. Boonman, Astron. Astrophys. 389, 446 (2002).


\end{document}